\documentstyle[aps,multicol,tabularx,rotate,epsf]{revtex}
\begin{document}

\title{Nonlinear Mechanical Response of DNA due to Anisotropic Bending Elasticity}

\author{Farshid Mohammad-Rafiee and Ramin Golestanian}

\address{Institute for Advanced
Studies in Basic Sciences, Zanjan 45195-159, Iran}

\date{\today}

\maketitle
\begin{abstract}
The response of a short DNA segment to bending is studied, taking
into account the anisotropy in the bending rigidities caused by
the double-helical structure. It is shown that the anisotropy
introduces an effective nonlinear twist-bend coupling that can
lead to the formation of kinks and modulations in the curvature
and/or in the twist, depending on the values of the elastic
constants and the imposed deflection angle. The typical wavelength
for the modulations, or the distance between the neighboring kinks
is found to be set by half of the DNA pitch.

\medskip
\noindent Pacs numbers: 87.15.-v, 36.20.-r, 61.41.+e
\end{abstract}

\begin{multicols}{2}

Genetic information in living cells is carried in the linear
sequence of nucleotides in DNA. Each molecule of DNA contains two
complementary strands of nucleotides that are matched by the
hydrogen bonds of the G-C and A-T base pairs (bps). The DNA
double-helix can be found in several forms that differ from each
in the geometrical characteristics such as diameter and
handedness. Under normal physiological conditions DNA adopts the
B-form, in which it consists of two helically twisted
sugar-phosphate backbones with a diameter $2.4$ nm, which are
stuffed with base pairs. The helix is right-handed with $10$ bps
per turn, and the pitch of the helix is $3.4$ nm. While
understanding some aspects of DNA functioning requires the details
of the chemical bindings etc., there are quite a number of
processes in cell that involve major structural and conformational
changes in DNA, where typically a small segment of DNA interacts
with certain proteins \cite{alb}.

For example, consider the transcription reaction catalyzed by RNA
polymerase, which creates an RNA strand that is complementary to
the transcribed DNA strand. The catalysis takes place inside the
transcription bubble: this is an $18$ bp stretch of unwound DNA
helix stabilized by RNA polymerase, which harbors a $12$ bp
RNA-DNA hybrid \cite{Stryer}. Low resolution structure of {\em E.
coli} RNA polymerase has shown a thumb-like projection that
seemingly closes around the DNA, and bends it to about
$90^\circ$-$120^\circ$ \cite{Darst}. There are also similar
visualizations of DNA polymerase, suggesting that the enzyme bends
a piece of DNA inside it \cite{Beese}. In nucleosomes, a $150$ bp
long segment of DNA wraps about $1.6$ turns around the histone
complex that has a diameter of about $11$ nm \cite{Luger}. In all
of these DNA-protein complexes, a DNA fragment is subjected to a
bending of the order of about $4^\circ$-$7^\circ$ per base pair,
which is typically a large curvature.

The idea of studying the response of DNA to mechanical stresses is
as old as the discovery of the double-helix structure itself
\cite{Wilkins}. An elastic description of DNA has been developed
over the past 20 years, taking on different approaches that
include Lagrangian mechanics \cite{Benham,Rudnick}, statistical
mechanics \cite{Marko1}, and molecular dynamics simulation
\cite{Schlick}. Motivated by the aforementioned structures in
various DNA-protein complexes, we study the bending of short DNA
segments using an elastic model. We assume that the length-scale
set by the helix pitch $P\simeq 3.4 \;{\rm nm}$ that involves $10$
base pairs, is long enough so that we can use a continuum elastic
theory, and yet short enough that the spatial anisotropy due to
the double-helical structure of DNA matters. The elastic model of
DNA is used with anisotropic bending rigidities, while neglecting
the linear phenomenological twist-bend \cite{Marko2} and
twist-stretch \cite{Kamien} couplings. Whereas an isotropic
elastic model predicts uniform bending and twisting, we find that
the anisotropy in the bending rigidity introduces an effective
nonlinear twist-bend coupling \cite{Golestanian,tannie} that could
lead to a variety of bending and twisting structures.

Depending on the relative strengths of the elastic constants and
the overall amount of deflection, we observe six different
regimes: (1) A regime in which curvature is localized in the form
of kinks in a periodic arrangement accompanied by simultaneous
suppression of twist, while the kinks are separated by virtually
flat (rod-like) and uniformly twisted segments, (2) a regime with
synchronized smooth modulations of the curvature and the twist,
where a rise in curvature corresponds to suppression of twist and
vice versa, (3) a regime with uniform curvature and modulated
twist, (4) a regime with the kink-rod structure in the curvature
and uniform twist, (5) a regime with modulated curvature and
uniform twist, and (6) a regime with both uniform curvature and
uniform twist. These different ``forms'' of bent DNA are presented
in Fig. 1 below, and in Fig. 2 it is specified for what values of
the parameters each form appears. We find that the period of
modulations as well as the distance between neighboring kinks is
always set by half of the DNA pitch, i.e. 5 bps, to a good
approximation. For a general DNA segment bending is always found
to reduce the linking number due to the effective (nonlinear)
twist-bend coupling. When the length of the DNA segment is equal
to an integer multiple of 5 bps, however, we observe a transition
from an over-twisted regime where there are kinks located exactly
at the end points to an under-twisted regime where the kinks are
located exclusively in the interior. The boundary corresponding to
this transition in the parameter space of the elastic constants is
plotted in Fig. 3.

The elastic model of DNA represents the molecule as a slender
cylindrical elastic rod. The rod is parametrized by the arclength
$s$, and at each point, an orthonormal basis is defined with
$\hat{e}_1$, $\hat{e}_2$, and $\hat{e}_3$, corresponding to the
principal axes of the elastic rod. In the undeformed state, all
the systems are parallel, with the $\hat{e}_3$-axes being all
parallel to the axis of the rod. The deformation of the elastic
rod is then parametrized by a mapping that relates the local
coordinate system to a reference one, specified by the Euler
angles $\theta(s)$, $\phi(s)$, and $\psi(s)$. The elastic energy
is written as \cite{Landau}
\begin{equation}
\beta E=\frac{1}{2}\int_0^L d s \;
[A_1\Omega_1^2+A_2\Omega_2^2+C(\Omega_3-\omega_0)^2] ,\label{E1}
\end{equation}
in the dimensionless units, where $A_1$ and $A_2$ are the bending
rigidities, $C$ is the twist rigidity, and $\omega_0=2\pi/P=1.85
\; {\rm nm}^{-1}$ is the spontaneous twist of the helix. In terms
of Euler angles, we have
$\Omega_1(s)=(d\theta/ds)\sin\psi-(d\phi/ds)\sin\theta\cos\psi$,
$\Omega_2(s)=(d\theta/ds)\cos\psi+(d\phi/ds)\sin\theta\sin\psi$,
and $\Omega_3(s)=(d\psi/ds)+(d\phi/ds)\cos\theta$. Without loss of
generality, we assume that the rod will stay planar even after the
application of a bending force, and hence set $d\phi/ds=0$. The
bending is characterized by an overall angle $\theta_0$, and we
assume that the bent segment of DNA is attached to undeformed
tails, so that it corresponds to bending of a piece of a long DNA.
Since the tails have a uniform twist characterized by $\omega_0$,
we assume that the twist at the two ends of the DNA segment is
also $\omega_0$, to preserve continuity.

After defining $A\equiv\frac{1}{2}(A_1+A_2)$,
$A^\prime\equiv\frac{1}{2}(A_2-A_1)$, and setting $s=L t$, we have
\begin{eqnarray}
\beta E = \frac{A^\prime\theta_0^2}{2L}\int_0^1 d t
&&\left[\left(\frac{A}{A^\prime}+\cos 2\psi\right)
\frac{\dot\theta^2}{\theta_0^2}\right.\nonumber \\
&&\left. +\frac{C}{A^\prime\theta_0^2}\left(\dot\psi- \omega_0 L
\right)^2\right],\label{E2}
\end{eqnarray}
where we have used $\dot\theta\equiv\frac{d\theta}{dt}$ and
$\dot\psi\equiv\frac{d\psi}{dt}$. The above equation shows
manifestly that there are three effective dimensionless parameters
in the system: $A/A^\prime$, $C/(A^\prime\theta_0^2)$, and
$\omega_0L$. We can find the extremum of $\beta E$ by applying
standard variational techniques to Eq. (\ref{E2}). The resulting
Euler-Lagrange equations for $\theta(t)$ and $\psi(t)$ are found
as
\begin{equation}
\frac{\ddot \theta }{\theta_0}=\frac {2\dot \psi (\dot
\theta/\theta_0) \sin2\psi}{\left(\frac{A}{A^\prime}+\cos
2\psi\right)},\label{D1}
\end{equation}
and
\begin{equation}
\ddot \psi=-\left(\frac{C}{A^\prime
\theta_0^2}\right)^{-1}\left(\frac{{\dot
\theta}^2}{\theta_0^2}\right) \sin2\psi.\label{D2}
\end{equation}
The above equations should be solved subject to the boundary
conditions $\dot \psi(0)=\dot \psi(1)=\omega_0 L$, $\theta(0)=0$
and $\theta (1)= \theta_0$. While the mean bending rigidity $A$ is
known to be about $50$ nm in the salt-saturation limit
\cite{Marko1}, and the twist rigidity $C$ is believed to be
roughly $75$ nm \cite{Vologodskii}, little is known about the
parameter $A^\prime$ that describes the anisotropy. Hence, we
solve the above set of coupled nonlinear equations numerically for
various values of the parameters $A/A^\prime$ and
$C/(A^\prime\theta_0^2)$, and for several DNA lengths.

\begin{figure}

\centerline{\epsfxsize=8.5cm\epsfbox{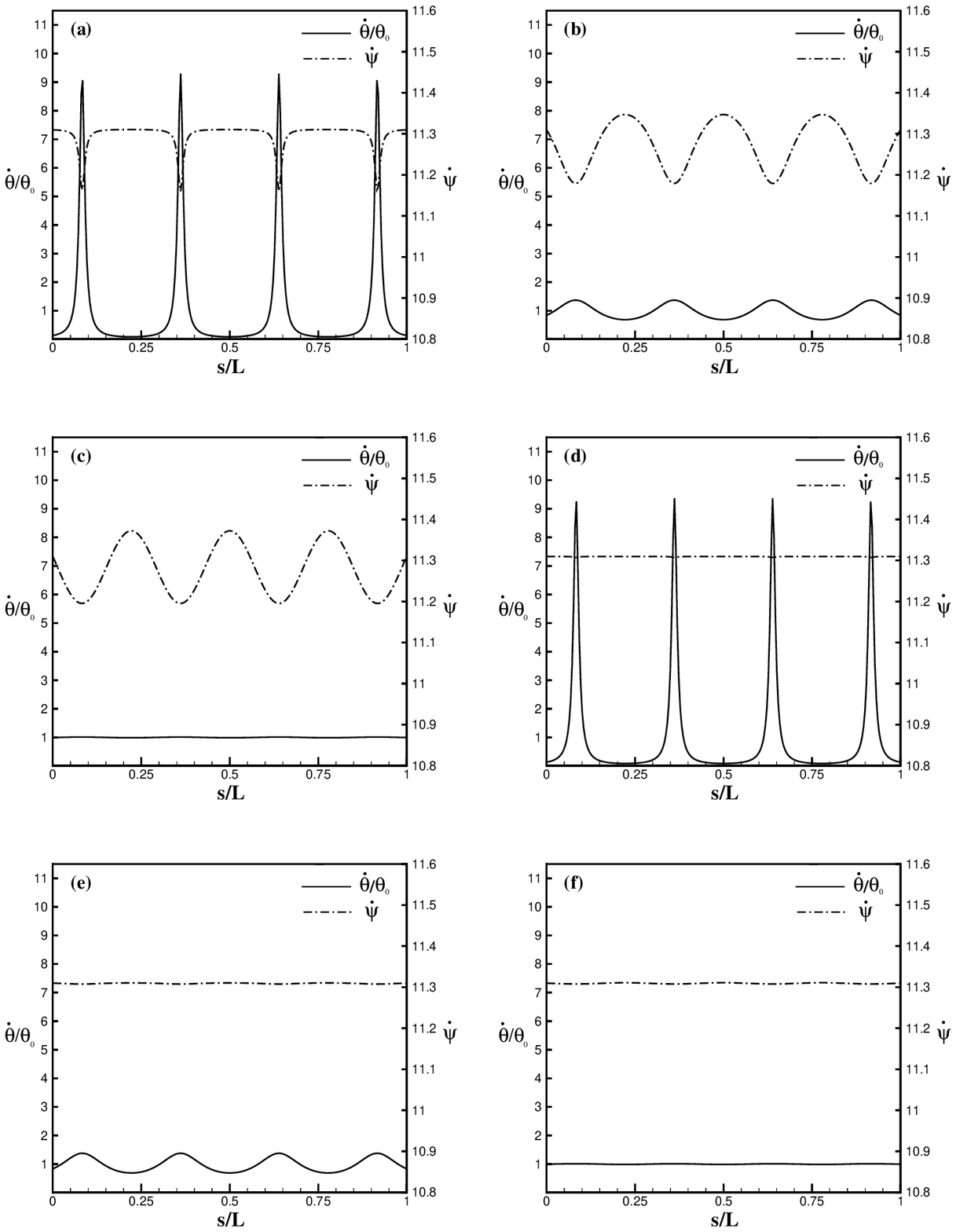}}

\caption{Different possible conformations of the DNA for various
values of $C/(A^\prime\theta_0^2)$ and $A/A^\prime$. The DNA
length is $18$ bps. The solid lines show the normalized curvature
$\dot{\theta}/\theta_0$ and the dotted lines show the twist
$\dot{\psi}$. (a) Kink-rod curvature structure and a corresponding
singular suppression of twist at the kinks. The plot corresponds
to $A/A^\prime=1.02$ and $C/(A^\prime\theta_0^2)=0.5$. (b)
Modulated curvature and twist. The plot corresponds to
$A/A^\prime=3$ and $C/(A^\prime\theta_0^2)=0.5$. (c) Uniform
curvature and modulated twist. The plot corresponds to
$A/A^\prime=79$ and $C/(A^\prime\theta_0^2)=0.5$. (d) Kink-rod
curvature structure and uniform twist. The plot corresponds to
$A/A^\prime=1.02$ and $C/(A^\prime\theta_0^2)=30$. (e) Modulated
curvature and uniform twist. The plot corresponds to
$A/A^\prime=3$ and $C/(A^\prime\theta_0^2)=30$. (f) Uniform
curvature and uniform twist. The plot corresponds to
$A/A^\prime=79$ and $C/(A^\prime\theta_0^2)=30$.}

\end{figure}

We observe that the above equations can acquire more than one
solution depending on the values of the parameters, with the
different DNA conformations characterized by deferent energies and
different values of the \textit{linking number}, defined as
$Lk=[\psi(1)-\psi(0)]/(2\pi)$ \cite{Lk}.  Note that for an
undeformed DNA we have $Lk_0=\omega_0 L/(2\pi)$. For each value of
DNA length that is not an integer multiple of 5 bps, we find six
distinct types of behaviors for various values of $A/A^\prime$ and
$C/(A^\prime\theta_0^2)$, as presented in Fig. 1 and described
above. For all of these conformations, which correspond to the
lowest energy case among the different possible solutions, the
change in the linking number $\Delta Lk=Lk-Lk_0$ is negative, and
DNA is thus under-twisted.

We also observe in all of these cases that the periodicity of the
curves, whether there are modulations or kinks, is set by half of
the DNA pitch, i.e. $5$ bps, to a good approximation. This can be
understood by looking at the first-integrals of Eqs. (\ref{D1})
and (\ref{D2}), which can be calculated as
\begin{equation}
\frac{\dot\theta}{\theta_0}=\frac{\alpha_1}{\left(\frac{A}{A^\prime}+\cos
2\psi\right)},\label{Tdot}
\end{equation}
and
\begin{equation}
\dot\psi^2=-\frac{\alpha_1^2}{\left(\frac{C}{A^\prime\theta_0^2}\right)
\left(\frac{A}{A^\prime}+\cos2\psi\right)}+\alpha_2,\label{psidot}
\end{equation}
respectively, where $\alpha_1$ and $\alpha_2$ are the integration
constants. Using a linear approximation $\psi \approx 2 \pi s/P$,
we can then expect a periodic behavior for both ${\dot \theta}$
and ${\dot \psi}$ in Eqs. (\ref{Tdot}) and (\ref{psidot}), with a
period of $P/2$.

\begin{figure}

\centerline{\epsfxsize=8.5cm\epsfbox{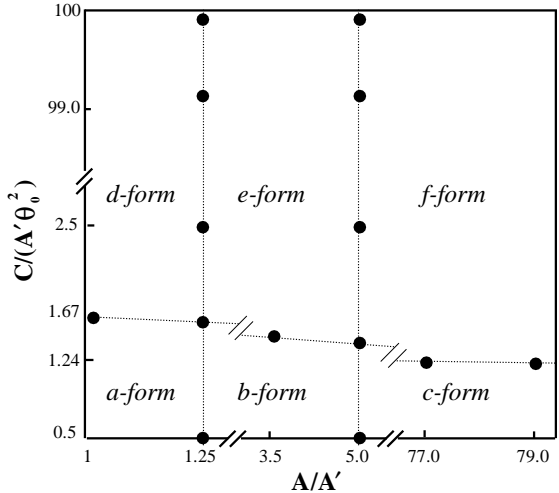}} \caption{A
diagram delineating the different regimes corresponding to the six
distinct forms of bent DNA, depending on the values of
$C/(A^\prime\theta_0^2)$ and $A/A^\prime$. The DNA length is $18$
bps. For fixed values of the elastic constants $A$, $A^\prime$,
and $C$, bending of DNA at various angles would correspond to
vertical lines. Note that in this case, the only permissible
transitions are from d-form to a-form, from e-form to b-form, and
from f-form to c-form, because the zone boundaries are also
vertical.}
\end{figure}

When the ratio $A/A^\prime$ (that is by definition always greater
than 1) is somewhat close to 1, the combination
$A/A^\prime+\cos2\psi$ could be close to zero, and then Eq.
(\ref{Tdot}) implies that $\dot\theta/\theta_0$ must be large.
This corresponds to the kinks in the curvature. In this case, for
sufficiently small $C/(A^\prime\theta_0^2)$ (numerically we find
$< 1.6$), Eq. (\ref{psidot}) implies that the same effect should
simultaneously happen in the twist, which corresponds to Fig. 1a.
For larger $C/(A^\prime\theta_0^2)$, however, this pattern should
disappear from the twist structure leaving only a uniform twist,
which corresponds to Fig. 1d.

For intermediate values of $A/A^\prime$, the periodicity imposed
by the term $A/A^\prime+\cos2\psi$ is still felt in the curvature,
in the form of modulations. Again, in this case, a small value for
$C/(A^\prime\theta_0^2)$ leads to accompanying modulations in the
twist, corresponding to Fig. 1b, while large
$C/(A^\prime\theta_0^2)$ will wash it out to uniform twist,
corresponding to Fig. 1e. Note that an increase in the curvature
always coincides with a decrease in the twist, and vice versa, as
can be manifestly understood from Eqs. (\ref{Tdot}) and
(\ref{psidot}).

When $A/A^\prime$ is very large, the dominant term in the
denominator of Eq. (\ref{Tdot}) is $A/A^\prime$, and hence the
curvature approaches a constant value. Since the parameters are
normalized such that a constant curvature should be equal to 1, we
find that in this limit $\alpha_1 \simeq A/A^\prime$. Then a small
value of $C/(A^\prime\theta_0^2)$ can make it possible for the
modulations to be felt in the twist, corresponding to Fig. 1c,
whereas for large values of $C/(A^\prime\theta_0^2)$ these
modulations are also washed out and we have a uniform twist, which
corresponds to Fig. 1f.

In Fig. 2, we have summarized the above results in a diagram that
delineates the different regimes corresponding to the different
forms of bent DNA defined in Fig. 1, for various values of
$C/(A^\prime\theta_0^2)$ and $A/A^\prime$. Note that while the
horizontal line (the boundary that separates a, b, and c regions
from d, e, and f regions) is slightly tilted, the other two lines
are virtually vertical within our numerical precision. Since for
fixed values of the elastic constants $A$, $A^\prime$, and $C$,
bending of DNA at various angles would correspond to vertical
lines, the fact that the two zone boundaries are vertical requires
that the only permissible transitions upon bending are from d-form
to a-form, from e-form to b-form, and from f-form to c-form. This
could provide an interesting possible experimental check of the
present analysis.

In the a-form regime, when the DNA length is between $5n$ bps and
$5(n+1)$ bps with $n$ being an integer number, the number of kinks
is $n+1$. One could then ask what happens if the DNA length is
exactly equal to $5n$ bps. In this case, two different sets of
solutions are observed, as shown in the insets of Fig. 3: (1) A
set of solutions in which there are $n$ kinks situated exclusively
in the interior, leaving an almost vanishing curvature at the
end-points. The value for $\Delta Lk$ in this case is negative,
which means that DNA is under-twisted. (2) Another set of
solutions in which there are $n-1$ kinks in the interior, and the
other kink is splitted in half and located exactly at the
end-points, leading to a large curvature at the two ends. In this
case, $\Delta Lk$ is positive, i.e. DNA is over-twisted. We found
that both of these sets of solutions could correspond to the
minimum energy case, depending on the values of the parameters.
The transition line between the two regimes is plotted in Fig. 3,
for a DNA segment of $15$ bps.

In the present treatment thermal fluctuations have been neglected.
This is justified by noting that the typical lengths considered
here are much smaller than the persistence length of DNA, which is
about $150$ bps, and therefore we do not expect considerable
deformation fluctuations of the DNA. However, due to the presence
of multiple energy minima that are not too different from each
other, we do expect that the conformation of DNA jumps between the
different possible minimum energy solutions, provided that the
barrier is not too large. This could be particularly interesting
if the bent DNA is involved in a dynamical process, such as
transcription \cite{farshid}.

\begin{figure}

\centerline{\epsfxsize=8.5cm\epsfbox{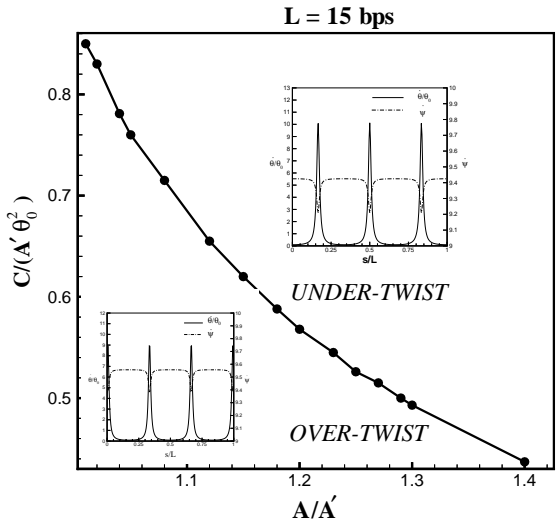}} \caption{A diagram
showing the different conformations of DNA, when its length is an
integer multiple of 5 bps. A transition is observed from an
over-twisted regime where there are kinks located exactly at the
end-points, to an under-twisted regime where the kinks are located
exclusively in the interior.}
\end{figure}

In conclusion, we have studied the response of a finite DNA
segment to bending, taking into account the anisotropy in the
bending rigidity caused by the double-helical structure. We have
shown that the nonlinear twist-bend coupling caused by the
anisotropy can lead to formation of kink-rod patterns and
modulations in the curvature and/or twist. This effect may be
related to the recent observation in high resolution X-ray, that
DNA is not uniformly bent around the histone octamer and its
curvature is larger at some places in the nucleosome \cite{Luger}.
We finally note that the present analysis can be also applied to
other biopolymers---the best candidate for an experimental
realization of these effects will probably be F-actin, which has a
much longer persistence length and can thus be more easily
manipulated in its stiff limit.


We are grateful to M.A. Jalali and T.B. Liverpool for helpful
discussions and comments.

\end{multicols}

\begin{references}
\bibitem{Lk}
Linking Number is a topological invariant for linear objects such
as the elastic rod, and according to White's theorem it can be
written as the sum of two terms: the twist and the writhe. In the
general case, we can write a simple expression for linking number
using the Euler-angle description as $Lk=(1/2\pi)\int_0^1 d t
\;(\dot\phi+\dot\psi)$ \cite{Rudnick}, which simplifies to
$[\psi(1)-\psi(0)]/(2\pi)$ in our case because $\dot\phi=0$.
\end{references}
\end{document}